\documentclass[12pt,twoside]{article}   
\usepackage[super,sort,comma]{natbib}

\usepackage{fancyhdr}		

\usepackage[section]{placeins}   %

\usepackage{graphicx}

\makeatletter \renewcommand\@biblabel[1]{$^{#1}$} \makeatother
\newcommand{\cen}[1]{\begin{center} #1 \end{center}}

\usepackage[mathlines]{lineno}

\usepackage{hyperref}
\hypersetup{ colorlinks,
	citecolor=blue,
	filecolor=blue,
	linkcolor=blue,
	urlcolor=blue
}

\usepackage{xcolor}

\definecolor{gray}{rgb}{0.6,0.6,0.6}
\definecolor{red}{rgb}{0.85,0,0}
\definecolor{green}{rgb}{0,0.85,0}
\definecolor{blue}{rgb}{0,0,0.85}
\definecolor{beige}{rgb}{0.92,0.87,0.78}
\definecolor{amaranth}{rgb}{0.9, 0.17, 0.31}
\definecolor{magenta}{rgb}{0.6, 0.4, 0.8}
\usepackage[all]{hypcap}    

\usepackage{amsmath}
\usepackage{amssymb}
\usepackage[normalem]{ulem}
\usepackage{upgreek}
\usepackage[title]{appendix}

\begin{document}

\cen{\sf {\Large {\bfseries First measurements of radon-220 diffusion in mice tumors, towards treatment planning in diffusing alpha-emitters radiation therapy} \\  
		\vspace*{10mm}
		Guy Heger$^{1,*}$, Mirta Duman\v{c}i\'{c}$^{1,*,\dag}$, Ishai Luz$^{2}$, Maayan Vatarescu$^{2}$, Noam Weizman$^{1,3}$, Brian W. Miller$^{4}$, Tomer Cooks$^{2,\ddag}$, Lior Arazi$^{1,\ddag}$} \\
  \vspace{5mm}
	$^1$Unit of Nuclear Engineering, Faculty of Engineering Sciences, Ben-Gurion University of the Negev, P.O.B. 653 Be'er-Sheva 8410501, Israel \\
     $^2$ The Shraga Segal Department of Microbiology, Immunology, and Genetics, Ben-Gurion University of the Negev, Be'er Sheva, Israel \\
     $^3$ Oncology Department, Radiation Therapy Unit, Hadassah – Hebrew University Medical Center, Israel \\
     $^4$ College of Medicine, Department of Radiation Oncology, Department of Medical Imaging, The University of Arizona, Tucson, AZ 85724 USA \\
     $^*$ These authors contributed equally \\
     $^\dag$ Current address: Gerald Bronfman Department of Oncology, Faculty of Medicine and Health Sciences, McGill University, Montreal, Quebec, Canada
      \vspace{5mm}\\
}

\pagenumbering{arabic}
\setcounter{page}{1}
\pagestyle{plain}
$^\ddag$ Authors to whom correspondence should be addressed. Lior Arazi: larazi@bgu.ac.il, Tomer Cooks: cooks@bgu.ac.il\\

\begin{abstract}
\noindent 
    {\bf Background:} Diffusing alpha-emitters radiation therapy (``Alpha-DaRT'') is a new method for treating solid tumors with alpha particles, relying on the release of the short-lived alpha-emitting daughter atoms of radium-224 from interstitial sources inserted into the tumor. Alpha-DaRT tumor dosimetry is governed by the spread of radium's progeny around the source, as described by an approximate framework called the ``diffusion-leakage model''. The most important model parameters are the diffusion lengths of radon-220 and lead-212, and their estimation is therefore essential for treatment planning.\\
    {\bf Purpose:} Previous works have provided initial estimates for the dominant diffusion length, by measuring the activity spread inside mice-borne tumors several days after the insertion of an Alpha-DaRT source. The measurements, taken when lead-212 was in secular equilibrium with radium-224, were interpreted as representing the lead-212 diffusion length. The aim of this work is to provide first experimental estimates for the diffusion length of radon-220, using a new methodology.\\
    {\bf Methods:} The diffusion length of radon-220 was estimated from autoradiography measurements of histological sections taken from 24 mice-borne subcutaneous tumors of five different types. Unlike previous studies, the source dwell time inside the tumor was limited to 30 minutes, to prevent the buildup of lead-212. To investigate the contribution of potential non-diffusive processes, experiments were done in two sets: fourteen \textit{in-vivo} tumors, where during the treatment the tumors were still carried by the mice with active blood supply, and ten \textit{ex-vivo} tumors, where the tumors were excised before source insertion and kept in a medium at $37^\circ$C with the source inside.\\
    {\bf Results:} The measured diffusion lengths of radon-220, extracted by fitting the recorded activity pattern up to 1.5~mm from the source, lie in the range $0.25-0.6$~mm, with no significant difference between the average values measured in \textit{in-vivo} and \textit{ex-vivo} tumors: $L_{Rn}^{in-vivo}=0.40\pm0.08$\,mm vs. $L_{Rn}^{ex-vivo}=0.39\pm0.07$\,mm. However, \textit{in-vivo} tumors display an enhanced spread of activity 2-3\,mm away from the source. This effect is not explained by the current model and is much less pronounced in \textit{ex-vivo} tumors.\\
    {\bf Conclusions:} The average measured radon-220 diffusion lengths in both \textit{in-vivo} and \textit{ex-vivo} tumors are consistent with published data on the diffusion length of radon in water and lie close to the upper limit of the previously estimated range of $0.2-0.4$~mm. The observation that close to the source there is no apparent difference between \textit{in-vivo} and \textit{ex-vivo} tumors, and the good agreement with the theoretical model in this region suggest that the spread of radon-220 is predominantly diffusive in this region. The departure from the model prediction in \textit{in-vivo} tumors at large radial distances may hint at potential vascular contribution, which will be the subject of future works.

\end{abstract}

%
\vspace{2pc}
\noindent{\it Running title}: Radon-220 diffusion measurement in mice tumors.\\
\newpage
%
%
%

\section{Introduction} \label{section:intro}

The radiobiological benefits of alpha particles in cancer treatment are well established \cite{hall2018radiobiology, Raju1991, MIRD_Pamphlet_22}. With a linear energy transfer (LET) of $60-230$ keV/$\upmu$m \cite{ASTAR}, they form dense tracks of ionization, which lead by direct interaction to irreparable complex DNA damage \cite{Goodhead1999,Hirayama2009}. At the peak ionization density of their tracks, the cell-killing efficacy of alpha particles is essentially unaffected by cellular oxygen levels, making them nearly as potent against hypoxic cells as against oxic ones \cite{Barendsen1966,Antonovic2013,Wozny2020}. Furthermore, the range of alpha particles in water, $\sim40-90\;\upmu$m, can in principle guarantee the sparing of healthy tissue. Given the promising therapeutic potential of alpha particles, multiple preclinical and clinical studies are underway towards their utilization in cancer treatment, as discussed in several recent reviews\cite{McDevitt2018,TAT_WG2018,Tafreshi2019}. Of these, $^{223}$RaCl${_2}$ treatments for castration-resistant prostate cancer with bone metastases are already commercial and in widespread clinical use \cite{Shirley2014}.

\begin{figure}[h]
    \centering
    \includegraphics[width=0.7\textwidth]{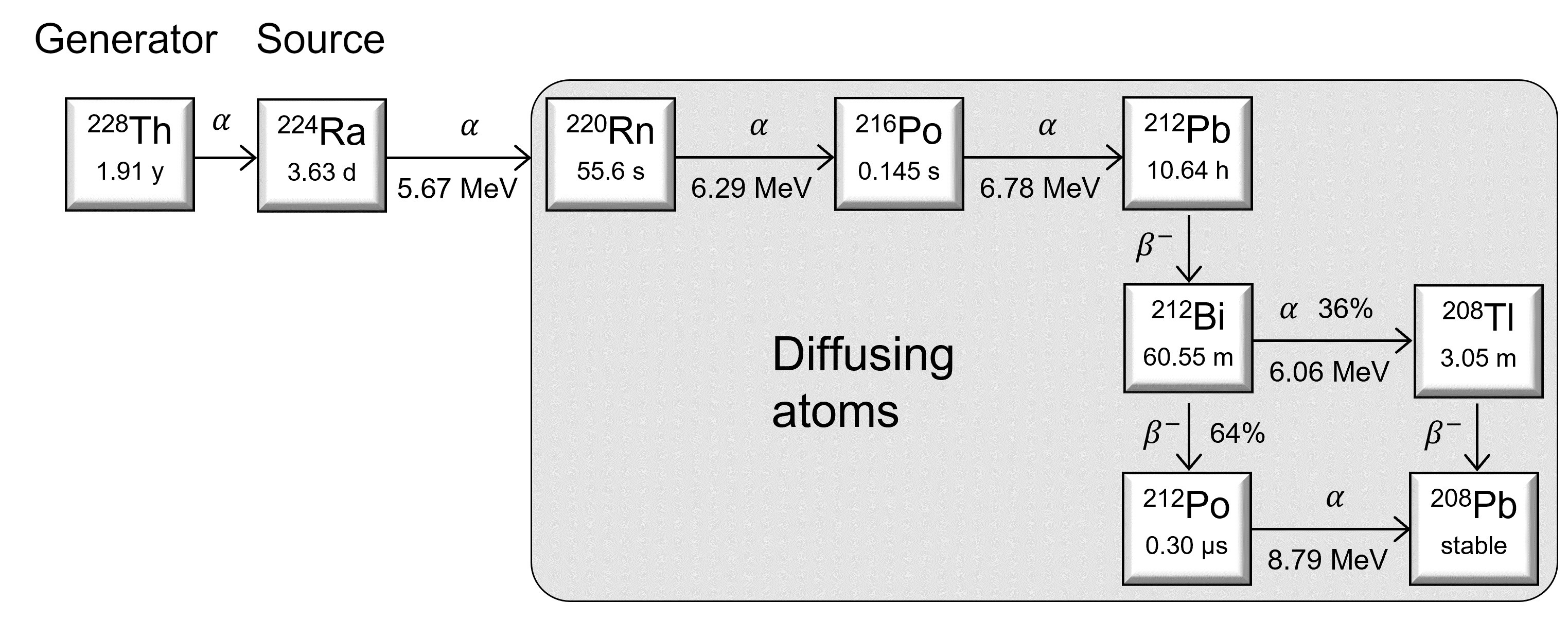}
    \caption{The $^{224}$Ra decay chain \cite{nudat}.}
\label{fig:RaDecayChain}
\end{figure}

Diffusing alpha-emitters radiation therapy (``Alpha DaRT'')\cite{Arazi2007}, enables the treatment of solid tumors using alpha particles. In Alpha DaRT, tumors are treated with sources carrying low activities of $^{224}$Ra below their surface. Once inside the tumor, the sources are designed to continuously release from their surface the short-lived daughter atoms of $^{224}$Ra: $^{220}$Rn, $^{216}$Po, $^{212}$Pb, $^{212}$Bi, $^{212}$Po and $^{208}$Tl. These spread by diffusion, with possible contribution from convective effects, creating---primarily through their alpha decays---a lethal high-dose region measuring several mm in diameter around each source. The full decay chain, along with the isotopes' half-lives, decay modes, and mean alpha particle energies is shown in Figure \ref{fig:RaDecayChain}.

Alpha DaRT was, and still is, extensively investigated in \textit{in vitro} and \textit{in vivo} preclinical studies on a large number of cancer types, as a stand-alone treatment \cite{Arazi2007, Cooks2008, Cooks2009a, Lazarov2011, Cooks2012}, in combination with chemotherapy \cite{Cooks2009b, Horev-Drori2012, Milrot2013, Reitkopf-Brodutch2015}, and as a stimulant of a local and systemic anti-tumor immune response \cite{Keisari2014, Confino2015, Confino2016, Domankevich2019, Domankevich2020, Keisari2020, Keisari2021, Del_Mare2023}. Since 2017, Alpha DaRT has been under clinical investigation, starting with locally advanced and recurrent squamous cell carcinoma (SCC) of the skin and head and neck \cite{Popovtzer2019}. The first-in-human trial, employed 2~$\mu$Ci $^{224}$Ra sources, inserted with a nominal spacing of 5~mm. The clinical outcomes of this trial showed promising efficacy and safety \cite{Yang2020}: all treated tumors (28/28) displayed positive response (shrank by $\sim30-100\%$), with 78.6\% (22/28) exhibiting complete response. Adverse effects were mild to moderate, with no observable local or systemic radiation-induced damage to healthy tissue. The alpha dose to all organs, resulting from $^{212}$Pb leaving the tumor through the blood, was calculated to be on the cGy level, with blood and urine activity measurements largely consistent with the prediction of a biokinetic model \cite{Arazi2010}. In one patient, untreated lesions shrank and disappeared when another was treated with Alpha DaRT (in the absence of any other treatment), suggesting an abscopal effect \cite{Bellia2019}. In a more recent study on recurrent or unresectable skin cancer, employing 3~$\mu$Ci sources at a nominal spacing of 4~mm, the complete response rate was 100\% (10/10) with only grade 1 and 2 adverse effects \cite{D’Andrea2023}.   

An in-depth description of the underlying physics of Alpha DaRT\cite{Arazi2020} introduced, as a first step towards treatment planning, a simplified theoretical approach: the ``Diffusion-Leakage (DL) model''. It was shown that of the six radionuclides released from the source, only the migration of $^{220}$Rn, $^{212}$Pb, and---to a lesser extent---$^{212}$Bi needs to be modeled, as their respective short-lived daughters are in local secular equilibrium. It was further shown that the spatial spread of the alpha dose around an Alpha-DaRT source is dictated by the \textit{diffusion lengths} of $^{220}$Rn and $^{212}$Pb---$L_{Rn}$ and $L_{Pb}$---which are the most critical parameters for Alpha-DaRT treatment planning. The theoretical and numerical work concerning the DL model was subsequently extended to realistic source geometries\cite{Heger2022a} and source lattices\cite{Heger2022b}, including a sensitivity analysis that considered changes in all model parameters, with an emphasis on the diffusion lengths. These calculations were later corroborated by independent works employing different finite-element methods \cite{Zhang2023, Cheve2023}. Alpha-dose calculations were recently augmented by a Monte-Carlo-based evaluation of the low-LET contribution in Alpha DaRT \cite{Epstein2023}.

Measurements of the spatial spread of $^{224}$Ra daughters in Alpha-DaRT-treated mice tumors\cite{Arazi2007, arazi_phd2009, Cooks2009a, Horev-Drori2012, Cooks2012, Reitkopf-Brodutch2015, Nishri2022} have so far been done $\sim4-5$~days after source insertion when $^{212}$Pb is in secular equilibrium with $^{224}$Ra, and the spatial activity profile has attained its asymptotic form\cite{Arazi2020}. Data analysis, done on a tumor-by-tumor basis by fitting autoradiography-based radial or cumulative $^{212}$Pb activity profiles \cite{arazi_phd2009}, or globally by fitting the effective diameter corresponding to specified dose levels as a function of source activity\cite{Arazi2020}, indicated---for SCC tumors---a mean value of $\sim0.6$~mm for the diffusion length, with individual measurements in the range $0.4-1.0$~mm\cite{arazi_phd2009}. This was interpreted as representing the diffusion length of $^{212}$Pb, while the diffusion length of $^{220}$Rn---which has never been measured directly---was estimated to lie in the range $\sim0.2-0.4$~mm, where the upper limit corresponds to its expected value in water\cite{Arazi2020, Jahne1987}. A limited number of measurements in pancreatic\cite{Horev-Drori2012}, melanoma\cite{Cooks2012}, prostate\cite{Cooks2012}, colon\cite{Cooks2012, Reitkopf-Brodutch2015}, and lung\cite{Cooks2009a} tumors in mice, indicated a spatial spread which was generally smaller than in SCC, with unpublished analysis indicating diffusion lengths in the range $\sim0.3-0.5$~mm. Recent autoradiography measurements of glioblastoma multiforme tumors in mice showed large variations in spread\cite{Nishri2022}, which---when analyzed in terms of diffusion lengths---corresponded to values ranging $\sim0.3-1$~mm.

The aim of this work is to provide, for the first time, experimental estimates of the diffusion length of $^{220}$Rn in mice tumors. We show that by limiting the source dwelling time inside the tumor to $\sim$30~minutes one can extract $L_{Rn}$ from the recorded activity profile. The study further aims to probe whether additional effects---in particular blood flow within the tumor---contribute to the recorded activity spread. This was done by implanting Alpha-DaRT sources in mice carcinoma before tumor excision (\textit{in vivo}) and after (\textit{ex vivo}). We find higher-than-expected values for $L_{Rn}$, as well as indications for non-diffusive effects that contribute to the alpha dose at large distances from the source \textit{in vivo} but not \textit{ex vivo}. We discuss the possible implications of these findings on the usability and limitations of the DL model for Alpha DaRT treatment planning.

\section{Methods}
\subsection{General procedure description}
Values of $L_{Rn}$ were extracted from autoradiography-measured sections of 24 mice-borne tumors. The data is divided into two sets: in the first, which contained 14 tumors and labeled \textit{in vivo}, a single Alpha-DaRT source was inserted into each tumor, still in the live animal, for 30 minutes. After 30 minutes (the reasoning for selecting this dwelling time is explained in section \ref{sec:Methodology}) the tumors were excised and the sources removed. The tumors were then set to freeze for 1 hour at -80$^{\circ}$C, and subsequently subjected to 10-$\upmu$m-thick histological sectioning (approximately at $90^{\circ}$ with respect to the source axis) by a LEICA CM 1520 cryostat. Finally, the sections were fixed with 4\% paraformaldehyde for 10 minutes and rinsed twice with PBS for 10 minutes each time. In the second set of experiments, which contained 10 tumors and labeled \textit{ex vivo}, the tumors were excised \textit{before} source insertion. These tumors (each with a single source inside) were then submerged in PBS and placed in an incubator for 30 minutes (37$^{\circ}$C, 5\% CO$_2$). The rest of the procedure was the same as for the \textit{in vivo} experiments (source removal, followed by freezing, sectioning and rinsing). The alpha particles emitted from the fixed tumor sections were imaged by the iQID camera \cite{iQID_paper} (see description below) and values of $L_{Rn}$ were then extracted from the resulting images.

\subsection{Source preparation}
Alpha-DaRT sources were produced by Alpha Tau Medical, following the same manufacturing methods used for clinical trials. The sources are made of stainless steel 316L cylindrical tubes, with an outer diameter of 0.7\;mm, inner diameter 0.4\;mm and length 6.5\;mm. Each source was loaded with $\sim$ 3 $\upmu$Ci $^{224}$Ra, firmly fixed to the source surface, with a $^{220}$Rn desorption probability\cite{Arazi2020} of $\sim45\%$.

\subsection{Cell cultures}
The study was conducted using SQ2 (SCC), 4T1 (triple-negative breast cancer), PANC2 (pancreatic duct adenocarcinoma), MC38 (colon carcinoma) and C4-2 (prostate carcinoma) cell lines. Due to a possible dependence of the diffusion lengths on the tumor microenvironment and tissue perfusion \cite{baker2011} that could further change with the tumor mass, the tumors were grown to various masses in the range of 0.13-1.6~g. The study was approved by the Ben-Gurion University Institutional Animal Care and Use Committee, and was conducted according to the Israeli Animal Welfare Act following the guidelines of the Guide for Care and Use of Laboratory Animals (National Research Council, 1996) [permits no. IL-78-12-2018(E) and IL-44-08-2021 (E)]. BALB/c, NOD scid gamma (NSG), and C57BL/6 mice were inoculated subcutaneously into the low lateral side of the back.

\subsection{Alpha-particle autoradiography}
For each tumor, several central histological sections were cut, usually $150-300$~$\upmu$m apart. The alpha particles emitted from the tumor sections were imaged by the iQID camera\cite{iQID_paper} for $15-60$ hours. A full description of the iQID system is given in Appendix \ref{appendix: iQID system}. Background values were calculated for each experiment individually by selecting a $4\times4$~mm$^2$ region clear of tumor sections, and averaging the number of alpha-particle hits per pixel in that region. The resulting values were subtracted from all section measurements of the corresponding experiment. The estimated source location was chosen, in most cases, according to the center of gravity (COG) of pixel activity in the section. In some sections where a visible dip was observed within the recorded high-activity region, the location was selected manually. An example of such a case is given in Figure \ref{fig:section_examples_dead}a, where the COG-estimated location is marked with a black dashed circle, while the manually chosen one is the blue circle. Next, the average number of alpha particle counts per pixel, $\langle N_\alpha (r)\rangle$, was calculated in concentric rings around the estimated source location at increasing radial distances. The width of the rings, $dr$, was set to 25.14 $\upmu$m, equal to the iQID's pixel size. The resulting ``number density profile'' was then used to estimate $L_{Rn}$ according to the procedure described in section \ref{sec:Methodology}. 

\subsection{Estimating the radon-220 diffusion length from measured number density profiles} \label{sec:Methodology}

Due to the very short half-live of $^{220}$Rn ($\sim1$~min) compared to the iQID imaging schedule after source removal (many hours), the recorded alpha decays from tumor sections are in fact those of $^{212}$Bi and $^{212}$Po, generated in the decay chain $^{220}$Rn$\rightarrow^{216}$Po$\rightarrow^{212}$Pb$\rightarrow^{212}$Bi$\rightarrow^{212}$Po. Since the measured tumors are frozen at -80$^{\circ}$C immediately after source removal and fixed with paraformaldehyde immediately after sectioning, redistribution of $^{212}$Pb and $^{212}$Bi relative to the decay locations of $^{220}$Rn is considered to be negligible. The reasoning for this relies on the known binding of $^{212}$Pb to large proteins\cite{Arazi2020} (promoting local fixation) and on experimental data indicating negligible redistribution of $^{212}$Bi relative to $^{220}$Rn and $^{212}$Pb even in live tumors at body temperature\cite{Arazi2020}. Therefore, one can write:

\begin{equation} \label{eq:activity_relation}
    N_{\alpha}(\textbf{r}) \propto \Gamma_{Bi}(\textbf{r}) \propto \Gamma_{Rn}(\textbf{r}),
\end{equation}
where $N_{\alpha}(\textbf{r})$ is the locally-recorded number of alpha particles per pixel, $\Gamma_{Bi}(\textbf{r})$ is the local specific activity of $^{212}$Bi when the iQID measurement begins, and $\Gamma_{Rn}(\textbf{r})$ is the local specific activity of $^{220}$Rn at the time of source removal from the tumor. Note that for the relevant specific activities ($<1$~Bq per pixel) the iQID's efficiency does not depend on the count rate \cite{iQID_paper}, ensuring linearity between $N_{\alpha}(\textbf{r})$ and $\Gamma_{Rn}(\textbf{r})$.

\paragraph{The theoretical expression of $\Gamma_{Bi}(r)$:} The expression for $\Gamma_{Bi}(r)$ is obtained by solving the DL model equations. (We write ``$r$'' rather than ``$\textbf{r}$'') because the calculation assumes that the activity is recorded in a plane perpendicular to the source axis and that there is axial symmetry.) While it would seem logical that for this, the DL equations should be solved for a realistic source geometry (a finite cylinder), in reality, it appears that choosing a finite line geometry better represents the actual measured data, as detailed further in the section \ref{section:results}. The expression of $\Gamma_{Bi}(r)$ for a finite line source is given below, with the reasoning behind this choice given at the end of the section:

\begin{equation} \label{eq:fit_function}
    \Gamma_{Bi}(r)=A\int_{-l/2}^{l/2} \frac{e^{-\sqrt{r^2+z^2}/L_{Rn}}}{\sqrt{r^2+z^2}} dz
\end{equation}

The integral is a sum of terms of the form $e^{-r'/L_{Rn}}/r'$ (the solution of the DL equations for a point source geometry \cite{Arazi2020}) with $r'=\sqrt{r^2+z^2}$, each representing a contribution of an infinitesimal segment $dz$ along the source length $l$, with $z=0$ at the source mid-plane, and $r$ the radial distance from the source in cylindrical coordinates.

\begin{figure}[!ht]
    \includegraphics[width=1\textwidth]{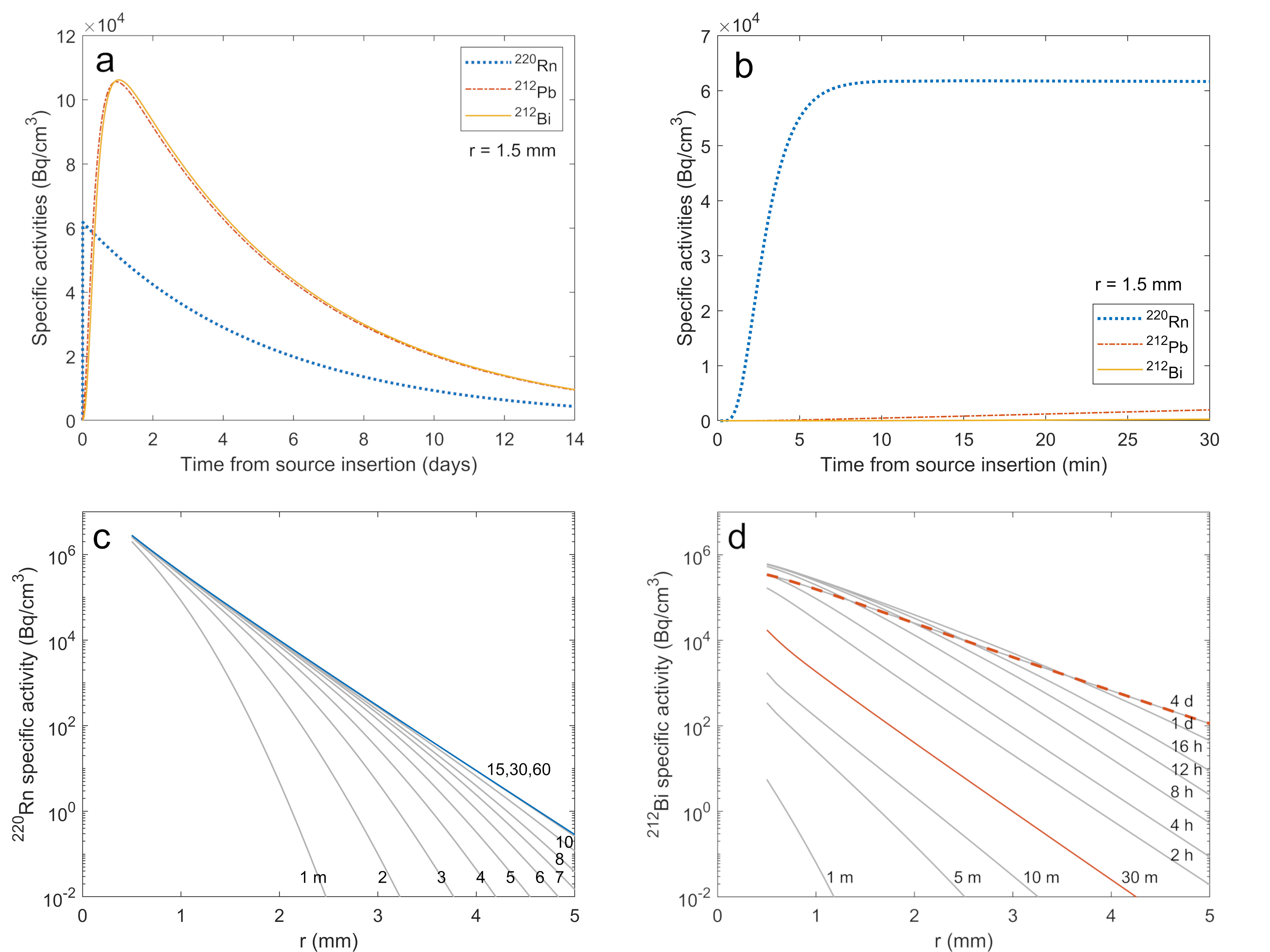}
    \caption{Complete time-dependent solution (specific activities) of the DL model equations for a 10-mm-long Alpha-DaRT line source, assuming $L_{Rn}=0.3$~mm and $L_{Pb}=0.6$~mm. The other model parameters are $P_{leak}(Pb)=0.6$, $L_{Bi}=0.1L_{Pb}$, $\alpha_{Bi}=0$. The specific activities are shown in the mid-plane of the source, which carries 3\;$\upmu$Ci $^{224}$Ra, with $P_{des}(Rn)=0.45$ and $P_{des}^{eff}(Pb)=0.55$. (a) The specific activities of $^{220}$Rn, $^{212}$Pb and $^{212}$Bi 1.5\;mm away from the source over 14 days; (b) Same as (a), focusing on the first 30 minutes; (c) Specific activity profiles of $^{220}$Rn over the first 60 minutes of source dwell time, with the blue curve corresponding to 30 minutes. The curves for 15 and 60 minutes overlap the blue one; (d) Same for $^{212}$Bi up to 4 days of source dwell time. The solid red curve denotes the profile at 30 minutes, and the dashed one at 4 days.}
\label{figure:Activities_vs_t_and_r}
\end{figure}

Each measured number profile $\langle N_\alpha (r)\rangle$ was fitted by the theoretical expression of $\Gamma_{Bi}(r)$ using least-squares optimization with $A$ and $L_{Rn}$ as free parameters. The fitted diffusion lengths $L_{Rn}$ (one for each section) were then averaged over the various sections of a given tumor for the final result.

\paragraph{Short vs.\ long source dwell times:} It is important to note that Equation (\ref{eq:fit_function}), with the slope governed by $L_{Rn}$, holds only for short source dwell times (the time between source insertion and removal) of up to a few hours. This is illustrated in Figure \ref{figure:Activities_vs_t_and_r}, which shows the time-dependent solutions of the DL model equations for the case $L_{Rn}=0.3$~mm, $L_{Pb}=0.6$~mm and $L_{Bi}=0.06$~mm. Panels (a) and (b) show the time-dependent specific activities of $^{220}$Rn, $^{212}$Pb, and $^{212}$Bi at $r=1.5$~mm as a function of time, considering long (days, panel a) and short (minutes, panel b) time scales. While the spatial buildup of $^{220}$Rn occurs over a few minutes, that of $^{212}$Pb and $^{212}$Bi takes days. In particular, as shown in (b), the specific activity of $^{212}$Bi is two orders of magnitude smaller than that of $^{220}$Rn 30 minutes after source insertion. Panel (c) shows a sequence of radial specific activity profiles of $^{220}$Rn, indicating that $\sim15$~min after source insertion, they stabilize on an asymptotic form over the entire relevant radial range. Panel (d) shows the same for $^{212}$Bi: at short time scales, the radial profile is exponential (linear on a logarithmic scale), with the same slope as that of $^{220}$Rn; over time the profile gradually changes, with a region of a milder slope (corresponding, here, to the larger diffusion length of $^{212}$Pb) expanding radially outward. Four days after source insertion, the radial profile of the specific activity of $^{212}$Bi becomes linear (on a log scale) and is completely dominated by $L_{Pb}$. As discussed further in Appendix \ref{appendix: closure calc}, the slope of the specific activity profile of $^{212}$Bi shows a two-step behavior as a function of time: a short-term step, with a plateau over $\sim10-100$~min after source insertion, corresponding to $L_{Rn}$, and a long-term step, reaching a plateau $\sim2000$~min ($\sim1.5$~d) later, corresponding to $L_{Pb}$. Therefore, autoradiography measurements performed $\sim4-5$~d after source insertion provide the dominant diffusion length $L_{dominant}=\textrm{max}(L_{Rn},L_{Pb})$, without being able to tell if it corresponds to that of $^{220}$Rn or $^{212}$Pb. On the other hand, the slope of the activity profile recorded $\sim30$~min after source insertion, necessarily corresponds to $L_{Rn}$. In the case $L_{Rn}>L_{Pb}$ the slope corresponds to $L_{Rn}$, and remains constant from $\sim10$~min after source insertion onward.

\paragraph{Fitting range:} An important choice in the fitting process is the range of $r$ over which the fit is performed. In the analysis of these experiments, two fit regions were used: $0.5-1$~mm for the \textit{in-vivo} set and $0.5-1.5$~mm for the \textit{ex-vivo} one. The reasoning behind these values is discussed in section \ref{section:results}. 

\paragraph{Comparing between \textit{in-vivo} and \textit{ex-vivo} experiments:} As the activity profiles measured in the \textit{in-vivo} sections showed significant departure from the model for $r\gtrsim1-1.5$~mm (while the \textit{ex-vivo} profiles followed the model more closely), a comparison parameter $\eta$ was defined as:

\begin{equation} \label{eq:eta}
    \eta(r) = \frac{\Gamma_{Measurement}(r)}{\Gamma_{Model}(r)}
\end{equation}
where $\Gamma_{Measurement}(r)$ is the measured activity profile of a given tumor section, and $\Gamma_{Model}(r)$ is the profile calculated using Equation (\ref{eq:fit_function}) with $L_{Rn}$ and $A$ obtained from the fitting process applied to the same section.

\paragraph{Validation using simulations:} Finally, simulations were performed with the aim of validating the approach described here. This was done by generating synthetic activity plots using known values of $L_{Rn}$, which were then ``blindly'' fitted with the theoretical expression with the aim of extracting $L_{Rn}$. The results of these calculations point to a convergence error of up to 3\% between the extracted value of $L_{Rn}$ using the fit process described here and the known values of $L_{Rn}$ used for the simulations. The full details of the validation process are given in Appendix \ref{appendix: closure calc}.

\begin{figure}[!ht]
    \centering
    \includegraphics[width=0.7\textwidth]{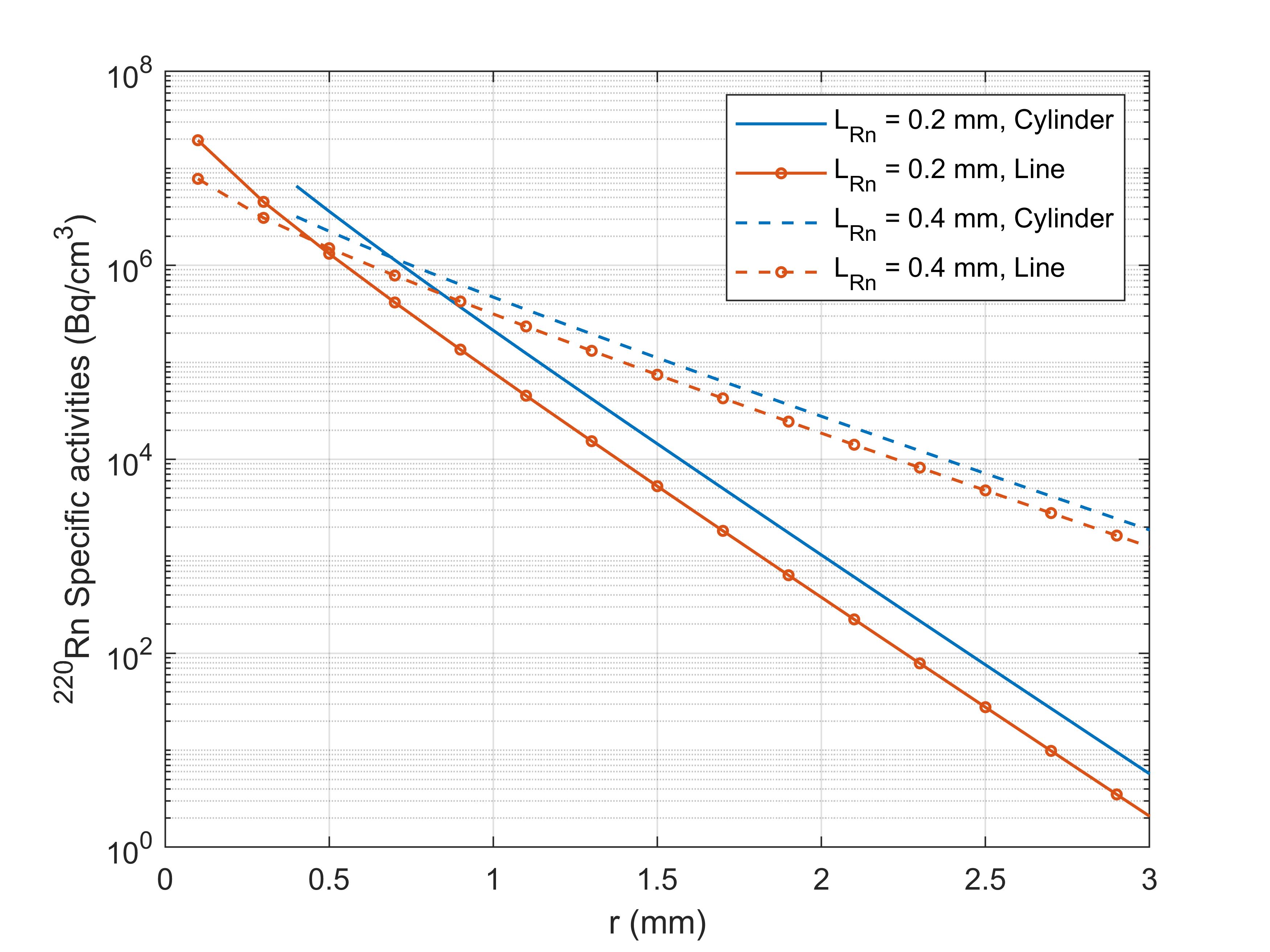}
    \caption{Comparison between 10-mm-long line and cylindrical Alpha-DaRT sources: $^{220}$Rn specific activity profiles, 30 min after source insertion, for two values of $L_{Rn}$.  The other model parameters are: $L_{Pb}=0.6$~mm, $P_{leak}(Pb)=0.6$, $L_{Bi}=0.1L_{Pb}$, $\alpha_{Bi}=0$. The specific activities are shown in the mid-plane of the source, which carries 3\;$\upmu$Ci $^{224}$Ra, with $P_{des}(Rn)=0.45$ and $P_{des}^{eff}(Pb)=0.55$.}
\label{figure:line_vs_source_somparison}
\end{figure}

\paragraph{Approximating the DaRT source as a finite line:} This choice is based on the measured activity profiles. As is shown in section \ref{section:results}, many of the profiles exhibit behavior more akin to a line source than a cylindrical one, i.\ e.\ the maximum of activity is at $r=0$ (or close to it) rather then the expected $r=0.35$ mm (the source radius). It could be that in these cases, the tumor tissue, which is still elastic after a short period of irradiation, shrinks radially back to fill the hole left by the source. In terms of the difference in expected activity profiles between the two cases (line vs. cylindrical source), the only change is that in the cylindrical case, the activity is ``pushed'' outward, but otherwise retains its spatial form. This can be seen in Figure \ref{figure:line_vs_source_somparison}, which shows activity profiles for both line and cylindrical sources for two values of $L_{Rn}$. This means that in the fitting process, the extracted diffusion length is not affected by the source radius (the free parameter $A$ does depend on it, but this parameter is of no interest in this study).

\subsection{Uncertainty analysis}
The uncertainties for extracted values of $L_{Rn}$ have three sources: (1) Selection of the source location (around which the activity profiles are calculated). Labeled $\Delta L_{Rn}^{loc}$, it was evaluated by comparing the values of $L_{Rn}$ extracted when the location was selected manually and the value, from the same section, when the location was selected by using the COG method. This comparison was done for 10 sections (five from \textit{in-vivo} tumors and five from \textit{ex-vivo} ones). The differences were then averaged, and the resulting uncertainty calculated to be 6.4\%. (2) Uncertainty introduced by the fitting process (discussed in section \ref{sec:Methodology} and shown in Appendix \ref{appendix: closure calc}, Figure \ref{figure:Fit_methodology}), labeled $\Delta L_{Rn}^{fit}$. Its value was taken as the maximal difference between the fitted and known values, as shown in Figure \ref{figure:Fit_methodology}b, resulting in an error of 3\%. (3) Variation between different sections obtained from the same tumor, labeled $\Delta L_{Rn}^{var}$. Its value, which was generally the dominant uncertainly term, was taken as the standard deviation (SD) of the values extracted from sections of the same tumor. This is the dominant uncertainty and is generally of the order of $\sim$~10\%. Finally, all three were added in quadrature to give the total uncertainty for each extracted $L_{Rn}$:

\begin{equation}
     \Delta L_{Rn} = \sqrt{\Big(\Delta L_{Rn}^{loc}\Big)^2 + \Big(\Delta L_{Rn}^{fit}\Big)^2 + \Big(\Delta L_{Rn}^{var}\Big)^2}
\end{equation}

\section{Results} \label{section:results}
\begin{figure}[h]
    \centering
    \includegraphics[width=0.75\textwidth]{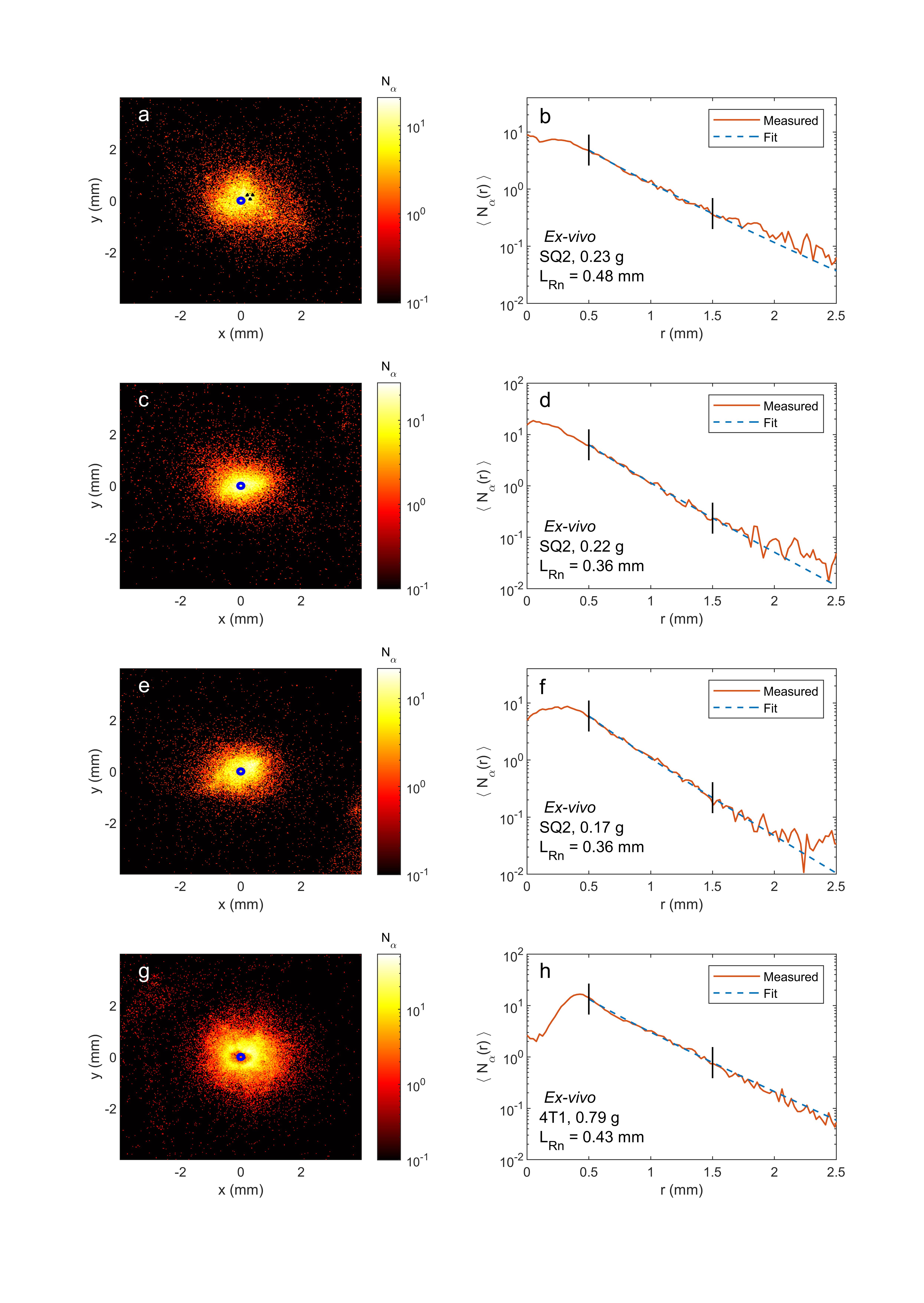}
    \caption{Examples of measured sections from four different experiments of the \textit{ex-vivo} tumors set. The left column shows the number of recorded alpha-particle hits per pixel for a single tumor section. In addition, it shows the estimated location of the source (blue circle). In panel (a), the location estimated using the COG method is also shown (black dots). The right column shows the ring-averaged number of alpha-particle hits per pixel as a function of distance from the source, as well as the fitted function used to extract $L_{Rn}$, Equation (\ref{eq:fit_function}). The vertical bars in the right-side panels show the start and end of the fit region ($0.5-1.5$~mm). Each row shows data from a different tumor; the specific cell line, tumor mass, and fitted $L_{Rn}$ are stated in the right-side panels.}
    \label{fig:section_examples_dead}
\end{figure}

\begin{figure}[h]
    \centering
    \includegraphics[width=0.75\textwidth]{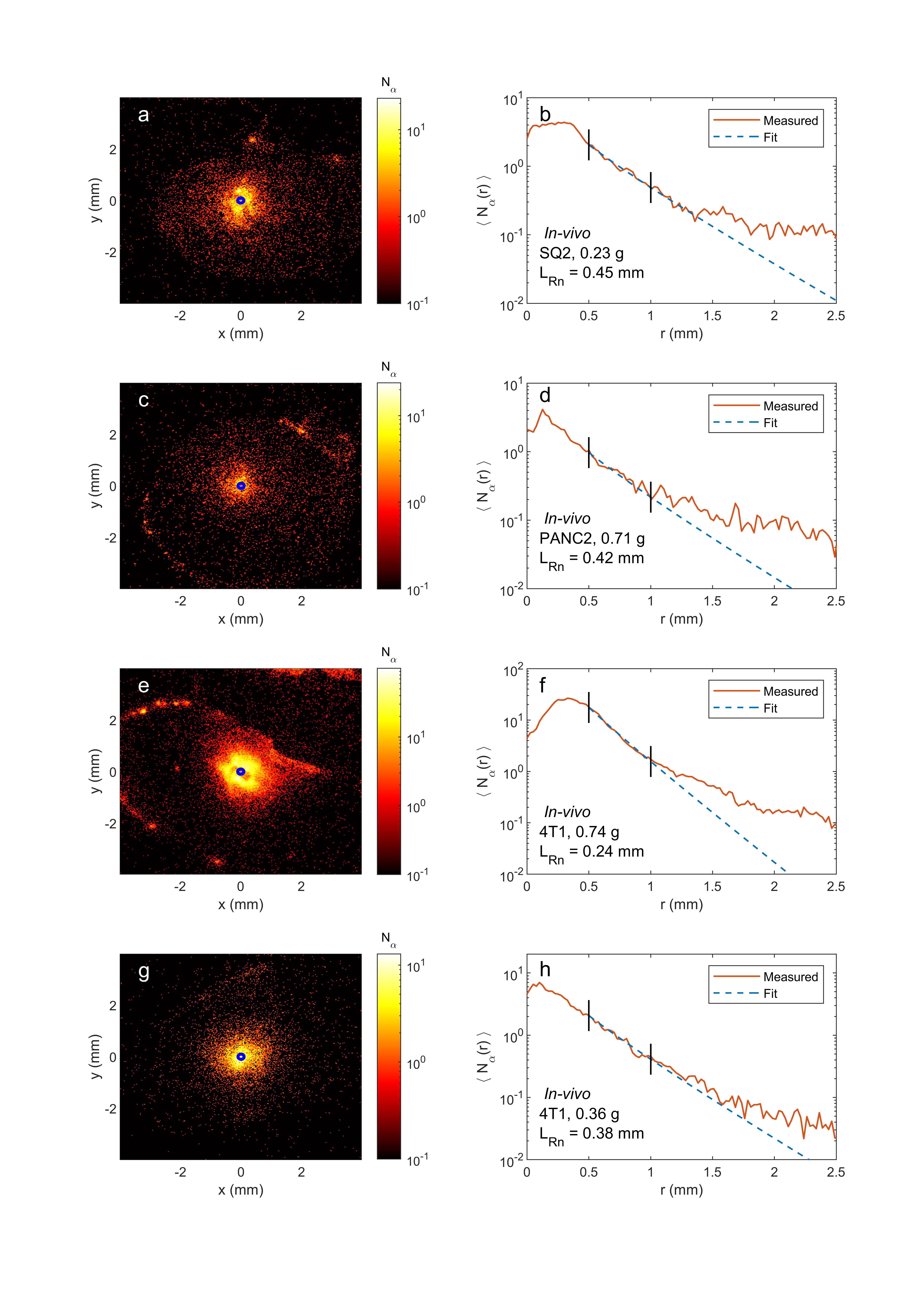}
    \caption{Examples of measured sections from four different experiments of the \textit{in-vivo} tumors set. The left column shows the number of recorded alpha-particle hits per pixel for a single tumor section. In addition, it shows the estimated location of the source (blue circle). The right column shows the ring-averaged number of alpha-particle hits per pixel as a function of distance from the source, as well as the fitted function used to extract $L_{Rn}$, Equation (\ref{eq:fit_function}). The vertical bars in the right-side panels show the start and end of the fit region ($0.5-1$ mm). Each row shows data from a different tumor; the specific cell line, tumor mass, and fitted $L_{Rn}$ are stated in the right-side panels.}
    \label{fig:section_examples_live}
\end{figure}

Figures \ref{fig:section_examples_dead} and \ref{fig:section_examples_live} show examples of measured sections from the \textit{ex-vivo} and the \textit{in-vivo} tumors sets, respectively (four different tumors in each figure). The left column shows the number of recorded alpha-particle hits per pixel $N_{\alpha}(\textbf{r})$ for a single tumor section. In addition, it shows the estimated location of the source (blue circle). In panel (a) of Figure \ref{fig:section_examples_dead}, the location estimated using the COG method is also shown (black dots), emphasizing the need for a manual selection of the source location in some cases. The right column shows the ring-averaged number of alpha-particle hits per pixel $\langle N_\alpha (r)\rangle$ as a function of distance from the source, as well as the fitted function used to extract $L_{Rn}$, Equation (\ref{eq:fit_function}). Although Equation (\ref{eq:fit_function}) refers to the specific activity, rather than to $\langle N_\alpha (r)\rangle$, the difference between the two is incorporated in the scaling factor $A$ which is a free parameter in the fit. Each row shows data from a different tumor: Figure \ref{fig:section_examples_dead}, panels (a)-(b): SQ2, 0.23 g, (c)-(d): SQ2, 0.22 g, (e)-(f): SQ2, 0.17 g, (g)-(h): 4T1, 0.79 g. Figure \ref{fig:section_examples_live}, panels (a)-(b): SQ2, 0.23 g, (c)-(d): PANC2, 0.71 g, (e)-(f): 4T1, 0.74 g, (g)-(h): 4T1, 0.36 g. The vertical bars in the right-side panels show the start and end of the fit region: 0.5 - 1 mm for \textit{in-vivo} tumors and 0.5 - 1.5 mm for \textit{ex-vivo} ones. The reason for the change of fit region is explained below. A good agreement between the measured activity profiles shown here and the theoretical ones, fitted using Equation (\ref{eq:fit_function}), indicates that indeed the fitting method, even though it is to an approximate source geometry (a superposition of point sources), ``captures'' the leading effects of physical processes at short radial distances (up to 1-1.5 mm from the source), namely, the diffusion of $^{220}$Rn.

Averaging over all diffusion length values extracted from the 14 \textit{in-vivo} tumors and comparing it to the 10 \textit{ex-vivo} ones, there seems to be no significant difference between the groups: 0.40 $\pm$ 0.08 mm for \textit{in-vivo} tumors and 0.39 $\pm$ 0.07 for \textit{ex-vivo} ones. However, looking at the dose profiles of the \textit{in-vivo} tumors (Figure \ref{fig:section_examples_live}), it can be seen that at larger radial distances the slope appears to change (decrease), an effect which is not observed in \textit{ex-vivo} tumor profiles (Figure \ref{fig:section_examples_dead}). This is the reason for the different fit regions between \textit{ex-vivo} and \textit{in-vivo} tumors.

\begin{figure}[h]
    \centering
    \includegraphics[width=0.7\textwidth]{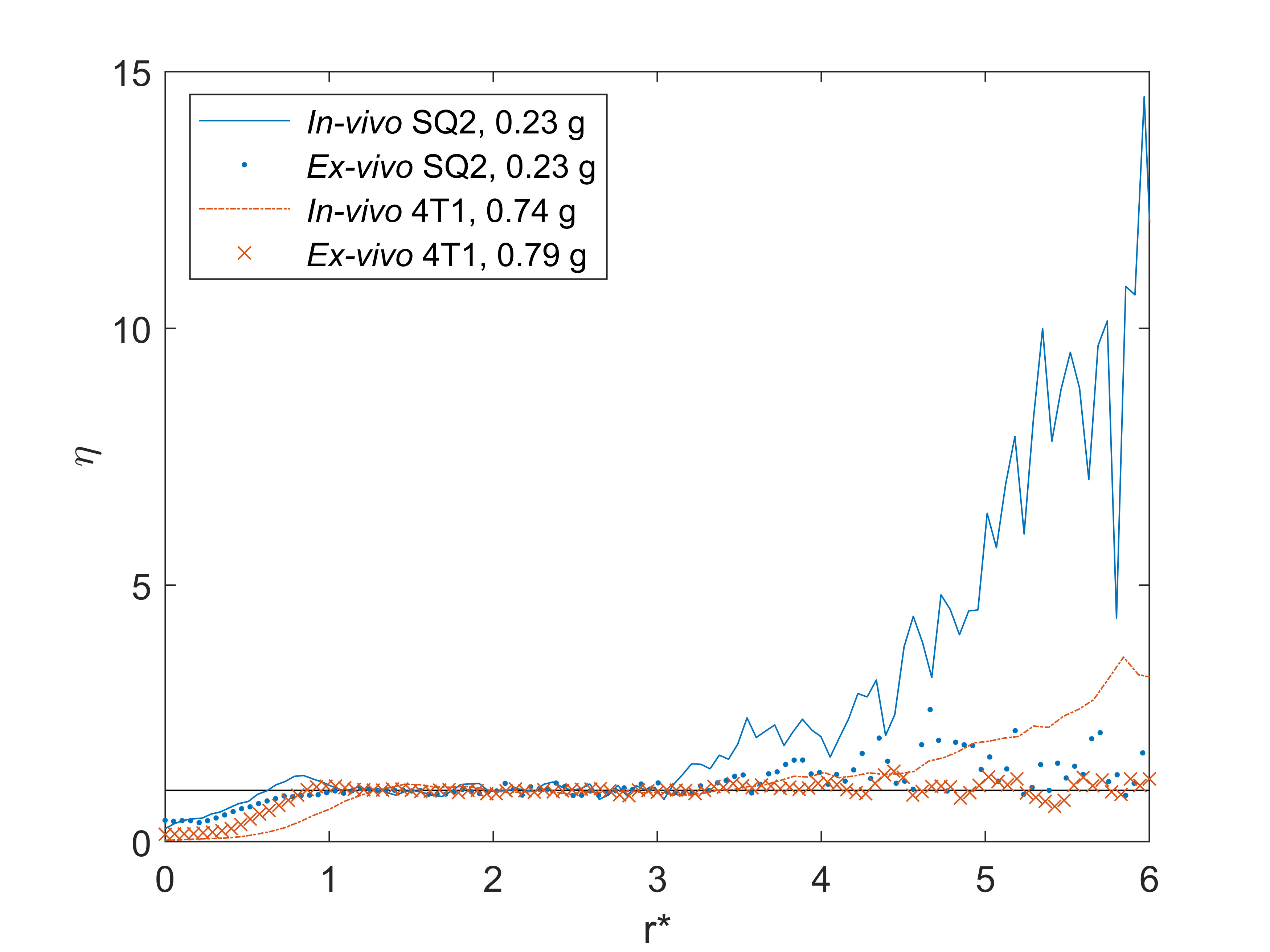}     
    \caption{Examples of calculated $\eta$ values as a function of distance from the source in two \textit{in-vivo} and two \textit{ex-vivo} tumors, expressed in units of diffusion length, $r^*=r/L_{Rn}$. The experiment set, cell line, and tumor mass are indicated in the figure. A horizontal line is added at $\eta=1$ for reference.}
    \label{fig:eta_vecs}
\end{figure}

\begin{figure}[h]
    \centering
    \includegraphics[width=1\textwidth]{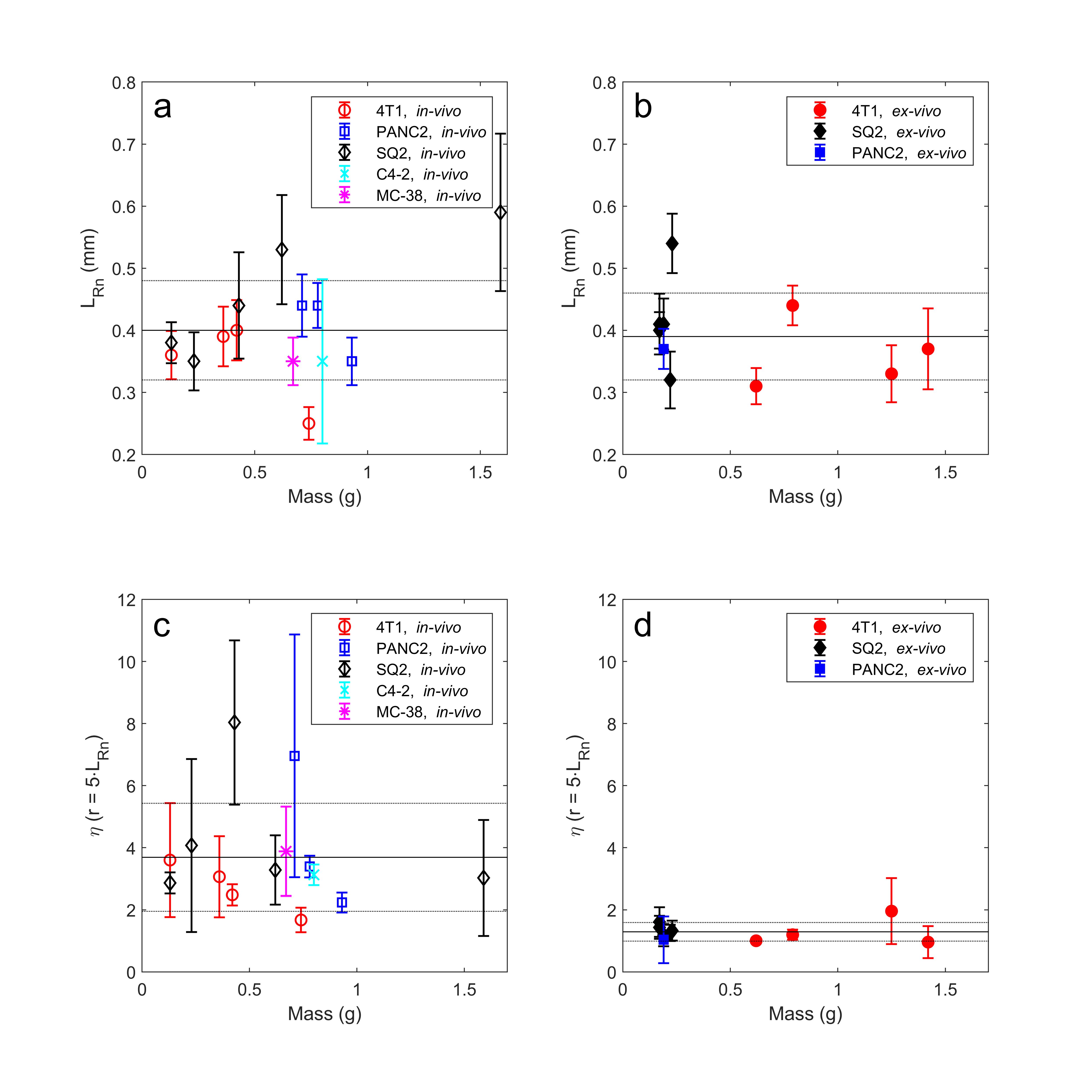}     
    \caption{(a) Experimental results for $L_{Rn}$ measured in 14 \textit{in-vivo} tumors. The horizontal line represents the average value of $L_{Rn}$ and the dotted lines are one standard deviation above and below that average. (b) The same, for the 10 \textit{ex-vivo} tumors. (c) Calculated $\eta$ values for the \textit{in-vivo} tumors, at distances equal to 5 times the measured diffusion length of the respective tumor. As in (a), the horizontal line is the average and the dotted lines are $\pm$ one standard deviation. (d) The same, for the \textit{ex-vivo} tumors.}
    \label{fig:results}
\end{figure}

We use $\eta(r)$, introduced in section \ref{sec:Methodology} (equation \ref{eq:eta}), to quantify the difference between the measured activity profile and the theoretical one. Examples for calculated values of $\eta$ are shown in Figure \ref{fig:eta_vecs}. These were calculated for the four sections shown in Figure \ref{fig:section_examples_dead}a, \ref{fig:section_examples_dead}g, \ref{fig:section_examples_live}a and \ref{fig:section_examples_live}e. The values of $\eta$ in Figure \ref{fig:eta_vecs} are plotted against $r^*$, defined as $r^* = r/L_{Rn}$, with $L_{Rn}$ being the diffusion length extracted for the specific section. It can be seen that indeed $\eta$ values increase noticeably above 1 at distances of $r^*>2$ for the \textit{in-vivo} SQ2 tumor and $r^*>4$ for the \textit{in-vivo} 4T1 tumor. Based on Figure \ref{fig:eta_vecs}, we chose $r^* = 5$ as a fixed reference point to evaluate the $\eta$, representing a point where significant deviation from the model exists while ensuring $\eta$ is calculated inside the tumor boundaries. The calculated $\eta(r^*=5)$ values are shown in Figure \ref{fig:results}a for the \textit{in-vivo} tumors and \ref{fig:results}b for the \textit{ex-vivo} ones. The horizontal line is the average value, and the dashed lines are $\pm$ one standard deviation. It can be seen that the values measured in \textit{in-vivo} tumors are significantly higher then those of \textit{ex-vivo} ones, suggesting that some contribution from non-diffusive processes, e.g., vascular flow exists, enhancing the spread of activity at large radial distances. While such processes are presently not included in the DL model, they have an apparently negligible effect close to the source, as indicated by the similar values of $L_{Rn}^{in-vivo}$ and $L_{Rn}^{ex-vivo}$ and by the good agreement with the model over the fit region.

The final results for the measured diffusion lengths are summarized in Figure \ref{fig:results}c for the \textit{in-vivo} tumors and \ref{fig:results}d for the \textit{ex-vivo} ones. As before, the horizontal line is the average value, and the dashed lines represent one standard deviation. For each tumor, at least two histological sections were analyzed.

\section{Discussion} \label{section:discussion}

The similar values of $L_{Rn}$ extracted from \textit{ex-vivo} and \textit{in-vivo} tumors and the good agreement with the model over the fit region $r=0.5-1$~mm or $r=0.5-1.5$~mm suggest that the dominant process governing the spread of activity over these regions is indeed diffusion (otherwise $L_{Rn}$ would be larger in \textit{in-vivo} tumors). A two-sample t-test comparing the \textit{in-vivo} and \textit{ex-vivo} $L_{Rn}$ values, including all 24 diffusion lengths measured (14 \textit{in-vivo} and 10 \textit{ex-vivo}) results in a \textit{p}-value of 0.73, showing indeed no statistically significant difference between the two groups. In addition, the average values in both \textit{in-vivo} and \textit{ex-vivo} tumors are consistent with the reported diffusion coefficient in water \cite{Jahne1987}, with some cases higher than water - even for \textit{ex-vivo} tumors (we note that other studies have also claimed higher values for $^{220}$Rn diffusion coefficient in water \cite{SINGH1993}). Previously in Alpha-DaRT \textit{in-vivo} studies, such values (measured above 0.4~mm) were associated with a contribution to diffusion from $^{212}$Pb, or $L_{Pb}$ \cite{Arazi2020} and led to the conclusion that, in most tumors, the spread of activity is ``lead-dominated'', i.e., the activity profile is dictated by the diffusion length of $^{212}$Pb and the dose delivered $\gtrsim1.5$~mm away from the source is the result of the single alpha particle emitted by either $^{212}$Bi or $^{212}$Po. Now, the high values of $L_{Rn}$ suggest that the Alpha-DaRT treatment may be, in fact, dominated by radon diffusion. This will be further discussed in future publications. It should be emphasized that the current $L_{Rn}$ results are  reported from a relatively small dataset. At this stage, it would be premature to conclude if $L_{Rn}$ varies across tumor types, as the variability within a given tumor type seems to be fairly large. Further measurements will be done to better quantify this.

It is important to address the enhanced spread of activity observed only in \textit{in-vivo} tumors (Figure \ref{fig:section_examples_live} and \ref{fig:results}c), which indicates that additional, beyond-diffusion processes, are at play. Interstitial flows should not significantly affect $^{220}$Rn spread: typical interstitial flow velocities are 0.1-1 $\upmu$m/s \cite{munson2014} so that even for an unrealistic uniform flow the convective length (the mean distance traveled by $^{220}$Rn before its decay). $L^{conv}_{Rn}=v\cdot\tau_{Rn}$ is $\sim8-80$ $\upmu$m, much smaller than the diffusion length. This process could affect small residual activities of $^{212}$Pb, where the much larger effective lifetime could lead to convective spread on the scale of the diffusion length or higher. A different process, in which vascularity could affect $^{220}$Rn spread, is a ``hop-on/hop-off'' mechanism, where $^{220}$Rn atoms enter a blood vessel, travel with the blood flow for a short distance, then leave the vessel and continue to spread via diffusion. A mechanism of this sort could potentially lead to a higher-than-zero leakage term for $^{220}$Rn (presumed negligible until now), but, as $^{220}$Rn is a noble gas, the ``transit time'' inside blood vessels is expected to be short, preventing significant leakage from the tumor. A recent study~\cite{Hinrichs2023} measuring radon solubility in various tissues also reported a lower solubility in diffusion-only experiments and \textit{in-vivo} samples where additional mechanism of distribution via blood flow is present~\cite{Nussbaum1957, Ishimori2017}. This, and other possible mechanisms, will be further investigated in future studies.

Finally, although the aim of this study is to provide, for the first time, an experimental value for $L_{Rn}$, we believe it is important to also relate these results to the Alpha-DaRT treatment, as these measurement are intended, ultimately, for use in treatment planning for Alpha DaRT. While the exact process leading to the enhanced spread is still unknown to us, this spread may, in principle, result in higher tumor doses than expected from the DL model using a conservative fit over short radial distances. However, for this to be the case the effect should be substantiated and shown to exist over a few days after source insertion. For now, our recommendation is to adopt the conservative approach and ignore it for tumor dosimetry. Such a conservative choice may be needed regardless of the additional non-diffusive spread, to compensate for unknown local variations in its magnitude. Even with such effects included, the dose fall-off in healthy tissue is sufficiently rapid to ensure sparing healthy tissue 2-3 mm away from the outermost source in a lattice. This will be shown in a separate publication where measurements in healthy pig tissues are discussed.

\section{Conclusions} \label{section:conclusions}

In this work we have presented the first measured values of the diffusion length of $^{220}$Rn, $L_{Rn}$, in both \textit{in-vivo} and \textit{ex-vivo} mice-borne tumors of varying types and masses. The measurement procedure, developed here for the first time, allows measurement of $L_{Rn}$ despite $^{220}$Rn's short half-life. The values measured for $L_{Rn}$ lie in the range of $0.25-0.6$~mm, which exceeds the previously assumed limit of 0.4~mm. This allows the scenario, presumed unlikely until now, that the dominant contributor to the spread of activity, in some tumors, is $^{220}$Rn, rather than $^{212}$Pb. This fact significantly increases the calculated dose profiles for an Alpha-DaRT treatment planning, easing the constraints on both source activities and lattice spacing. It was further shown that the measured activity profiles exhibit a pure-diffusion-based spread near the source surface ($r<1$ mm). However, an excess of activity was consistently observed in \textit{in-vivo} tumors far from the source, which cannot be explained by the current model. This will be the target of future studies.

\section*{Acknowledgements}
The SQ2 (squamous cell carcinoma), 4T1 (triple-negative breast cancer), PANC2 (pancreatic duct adenocarcinoma) and MC38 (colon carcinoma) cell lines were provided by Prof.\ Yona Keisari (Sackler School of Medicine, Tel-Aviv University, Israel). The C4-2 (prostate carcinoma) cell line was provided by Dr.\ Martina Bene\v{s}ov\'{a} (DKFZ-Heidelberg, German Cancer Research Center Foundation, Heidelberg, Germany).

\section*{Conflict of Interests}
This work was partly funded by Alpha Tau Medical Ltd (ATM). ATM was not involved in the study design and analysis, the writing of this article, or the decision to submit it for publication. L.A. and T.C. are minor shareholders of ATM. The scholarship of G.H. is partially paid by ATM and that of M.D. was fully paid by ATM through a research agreement with Ben-Gurion University of the Negev. The salary of I.L. is partially paid by ATM and that of M.V. is fully paid by ATM. The scholarship and salary of N.W. are partially paid by ATM. L.A. and T.C. are co-inventors of several Alpha-DaRT-related patents and patent applications. G.H., M.D., M.V., and I.L. are co-inventors of pending patent applications on Alpha-DaRT dose calculations.

\section*{Data availability}
The experimental data is available upon request to the corresponding author.

\begin{appendices}
\section{The iQID autoradiography system} \label{appendix: iQID system}
\paragraph{General description}
The activity present inside a tumor section was recorded using the ionizing-radiation Quantum Imaging Detector (iQID) autoradiography system \cite{iQID_paper}. This system consists of a scintillation material (a thin sheet of ZnS:Ag), an image intensifier, a series of optical lenses and a fast CMOS camera connected to a computer. During measurement, the tumor sections are placed on the scintillation foil. When an alpha particle hits the ZnS:Ag layer, a flash of light is created and amplified by the image intensifier (beta electron events are excluded by setting a high threshold on the light intensity). The light is then focused on the CMOS panel. Each flash of light recorded by the CMOS is referred to as a ``cluster''. The imaged clusters are continuously sent to the computer, where each cluster's central peak of intensity (i.e., its center-of-gravity, COG) is found. The cluster COG serves to estimate the emission point of the alpha particle. Given the 10\;$\upmu$m thickness of the tumor section, the accuracy in estimating the emission point is $\sim20$\;$\upmu$m and, since these are estimated on an event-by-event basis, there is no need to perform image deconvolution on the data. The iQID is usually operated at a refresh rate such that there is no overlap of clusters in a frame. In this study, it was operated at a refresh rate of 25 Hz. Finally, each cluster's COG coordinates in the $xy$ plane are saved in a list-mode file, along with its time-stamp.

\paragraph{iQID components}
The scintillation material used in this study is a 25 $\upmu$m layer of ZnS:Ag deposited on a 100 $\upmu$m layer of transparent polyester plastic sheet (Eljen technologies, EJ-440-100). The ZnS density is 3.25 $\pm$ 0.25 mg/cm$^2$. The iQID system used in this study is a commercial system from QScint Imaging Solutions, with a 50-mm-diameter active area and pixel resolution of 24.15 $\upmu$m. The device's efficiency was measured to be $0.46 \pm 0.04$ (coming from a one-sided measurement of the sample activity).

\paragraph{Efficiency calibration} The iQID's alpha particle detection efficiency was calculated using two $^{212}$Pb calibration samples. The samples were prepared by the following procedure. A small cylindrical plastic container (28.5~mm in diameter and 3~mm in height) is placed inside a low-pressure chamber ($<0.1$ mbar) with an open $^{228}$Th surface source at the bottom of the container and an aluminum sheet glued to its top. $^{224}$Ra ions recoiling from the $^{228}$Th source are implanted into the aluminum sheet. After several days of collecting $^{224}$Ra ions, the activity on the Al sheet approaches secular equilibrium with the parent $^{228}$Th source. In order to prepare a calibration sample, the $^{224}$Ra-implanted container top is removed and placed on top of another container, with a plastic sheet at the bottom which will become the calibration sample. The $^{224}$Ra-implanted Al sheet, similarly to what happens inside a tumor, releases its daughter atoms by recoil into the container and the calibration sample at the bottom. The exposure duration determines the calibration sample activity. Once this stage is finished, the calibration sample is taken to a NaI(Tl) gamma counter \cite{hidex} to determine its $^{212}$Pb activity. Finally, the sample is taken to the iQID system, where the alpha particles emitted from $^{212}$Bi and $^{212}$Po are measured. In order to compare the activity of $^{212}$Pb with that of $^{212}$Bi and $^{212}$Po, the calibration sample is set to ``rest'' for a few hours after collection, to ensure it reaches secular equilibrium. The two calibration sample activities used here were set to $\sim10$ and 20~Bq, and the resulting efficiency of the iQID was measured to be $0.46 \pm 0.04$.

\section{Fit methodology validation} \label{appendix: closure calc} 
The validity of the $L_{Rn}$ extraction process presented here was tested using the DART2D code \cite{Heger2022a}, which allows simulating the procedure of these experiments. These simulations produced artificial activity profiles for varying values of source dwell times and diffusion lengths, which were then blindly fitted using the process described in section \ref{sec:Methodology} to extract an ``unknown'' value of $L_{Rn}$, labeled $L_{fitted}$. These extracted values were then compared to the known parameters used in the simulation. 

The artificial activity profile were calculated as follows. DART2D \cite{Heger2022a} was used to solve the time-dependent DL model equations for $^{220}$Rn, $^{212}$Pb and $^{212}$Bi for a series of source dwell times ranging from minutes to days. For each case, the system was allowed to evolve uninterrupted from source insertion to the prescribed dwell time. The source was then ``switched off'' and the diffusion coefficients of $^{220}$Rn, $^{212}$Pb and $^{212}$Bi were set to zero (simulating negligible diffusion in the frozen---and then fixated---tumor tissue). The simulation continued further under these conditions, with the dynamics reduced to that of the local decay chain $^{220}$Rn $\rightarrow ^{212}$Pb $\rightarrow ^{212}$Bi. To simulate the autoradiography stage, the local number of $^{212}$Bi decays was integrated at the source mid-plane, for a duration consistent with the start and end times of the actual measurement (typically starting 3\;hours after tumor removal and lasting $\sim24$\;hours). The recorded $^{212}$Bi decays were then used to calculate the activity as a function of distance from the source (the activity profile) in the source mid-plane.

\begin{figure}[h]
    \centering
    \includegraphics[width=1\textwidth]{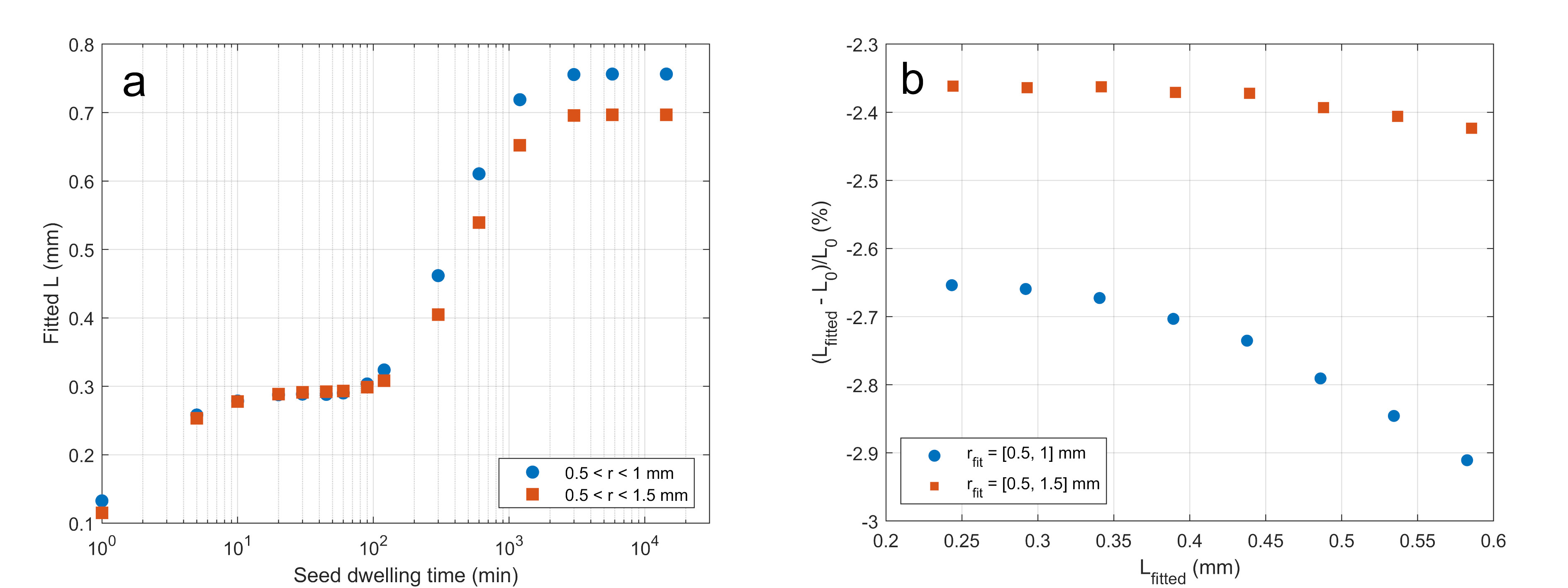}
    \caption{(a) Fitted values of $L$ depending on the source dwell time for two different radial regions: $0.5 < r < 1$ mm (blue circles) and $0.5 < r < 1.5$ mm (red squares). The diffusion lengths used for the simulated activity profiles are $L_{Rn}=0.3$~mm, $L_{Pb}=0.6$~mm. The source dwell time is 30 minutes and the leakage probability is $P_{leak}(Pb)=0.6$. The other model parameters are the same as before. (b) Relative error in the fitting process as a function of the fitted diffusion length value, for the two radial fitting ranges used in the data analysis.}
\label{figure:Fit_methodology}
\end{figure}

Figure \ref{figure:Fit_methodology}a shows the results (fitted diffusion length vs.\ dwell time) obtained for a case where $L_{Rn} = 0.3$ mm, $L_{Pb} = 0.6$ mm and the other model parameters are as before. The results are plotted for two different radial fit regions mentioned above. In both cases, the fit converged to the simulated value of $L_{Rn}$ for dwell times of several dozen minutes. On the other hand, when the dwell time is long (days), the fit converges to a value which is within $\sim10-25\%$ of the dominant diffusion length---here $L_{Pb}$---with an over-estimation than decreases when the fit is applied over a wider radial region. The accuracy of convergence for various values of $L_{Rn}$ was calculated as the relative error between the fitted value, $L_{fitted}$ and the one used for the calculation, $L_{0}$. The results, shown in figure \ref{figure:Fit_methodology}b for $L_{Pb}=0.6$~mm and a source dwell time of 30 minutes, point to a convergence error of up to 3\%. Based on these results, the experimental dwell time was set to 30 minutes, as a reasonable compromise allowing, on the one hand, sufficient time for the fit to approach $L_{Rn}$, but at the same time minimizing the buildup of $^{212}$Pb. We note that when the calculation was repeated for other values of $L_{Pb}$, namely $L_{Pb}=0.3$~mm and $L_{Pb}=0.1$~mm, both for $L_{Rn}=0.3$~mm, the 3\% convergence limit still held. Finally, we note that the offset along $z$ relative to the source midplane has a negligible effect ($<1\%$ change in convergence error), as long as the distance from the source edges is larger then $\sim$1 mm.\\

\end{appendices}

\section*{References}
\bibliography{bibli}
\bibliographystyle{./medphy.bst}

\end{document}